\documentclass[doublecol]{epl2}
\usepackage{amsmath,amssymb}
\usepackage{graphicx}

\newcommand{\dd}{\mathrm{d}}

\newcommand{\pd}[2]{\frac{\partial #1}{\partial #2}}
\newcommand{\fd}[2]{\frac{\delta #1}{\delta #2}}
\newcommand{\mean}[1]{\langle #1 \rangle}
\newcommand{\Int}[1]{\int\dd #1\;}
\newcommand{\IInt}[3]{\int_{#2}^{#3}\dd #1\;}

\renewcommand{\vec}[1]{\mathbf #1}

\newcommand{\x}{\vec r}         

\newcommand{\shr}{\dot\gam}     
\newcommand{\nois}{\boldsymbol\xi}
\newcommand{\id}{\mathbf 1}

\newcommand{\gam}{\gamma}
\newcommand{\eps}{\varepsilon}

\newcommand{\kap}{\kappa}

\newcommand{\kT}{k_\text{B}T}

\begin{document}


\title{Mobility and Diffusion of a Tagged Particle in a Driven Colloidal
  Suspension}

\author{Boris Lander\inst{1} \and Udo Seifert\inst{1} \and Thomas
  Speck\inst{2,3}}

\institute{
  \inst{1} {II.} Institut f\"ur Theoretische Physik,
  Universit\"at Stuttgart, Pfaffenwaldring 57, 70550 Stuttgart, Germany \\
  \inst{2} Department of Chemistry, University of California, Berkeley,
  California 94720, USA \\
  \inst{3} Chemical Sciences Division, Lawrence Berkeley National Laboratory,
  Berkeley, California 94720, USA
}

\abstract{We study numerically the influence of density and strain rate on the
  diffusion and mobility of a single tagged particle in a sheared colloidal
  suspension. We determine independently the time-dependent velocity
  autocorrelation functions and, through a novel method, the response
  functions with respect to a small force. While both the diffusion
  coefficient and the mobility depend on the strain rate the latter exhibits a
  rather weak dependency. Somewhat surprisingly, we find that the initial
  decay of response and correlation functions coincide, allowing for an
  interpretation in terms of an 'effective temperature'. Such a
  phenomenological effective temperature recovers the Einstein relation in
  nonequilibrium. We show that our data is well described by two expansions to
  lowest order in the strain rate.}


\maketitle


\section{Introduction}

The mobility of a single spherical particle immersed in a solvent determines
the velocity of the particle in response to an applied external force. For
small Reynolds numbers Stokes' law yields the famous expression
$\mu_0^{-1}=3\pi\eta a$ in terms of the sphere diameter $a$ and the solvent
viscosity $\eta$ in thermal equilibrium. This free mobility $\mu_0$ is
intimately related to spontaneous solvent fluctuations through the Einstein
relation. For a suspension of interacting particles, even without hydrodynamic
coupling, the mobility $\mu$ of a single tagged particle is reduced. This
reflects the fact that work is necessary to displace neighboring particles in
order for the tagged particle to move, leading to larger dissipation. Still,
in equilibrium the Einstein relation
\begin{equation}
  \label{eq:er}
  D = \kT\mu
\end{equation}
equates the effective, long-time diffusion coefficient $D$ obtained from
measuring the mean square displacement of a single tagged particle with its
mobility through the solvent temperature $T$, where $k_\text{B}$ is the
Boltzmann constant.

The Einstein relation~(\ref{eq:er}) is one out of many fluctuation-dissipation
relations valid in the \textit{linear response regime} for small perturbations
of the equilibrium state~\cite{kubo}. It is crucial to realize that also
nonequilibrium steady states allow for a linear response. However, driving the
suspension beyond the linear response regime, fluctuation-dissipation
relations such as the Einstein relation~\eqref{eq:er} need to be generalized
to nonequilibrium. There are two basic strategies discussed in the
literature. The first strategy is to introduce an additive correction taking
on the form of another correlation function~\cite{cris03,diez05,spec06}. This
correlation function involves another observable that can be related either to
entropy production~\cite{seif10} or 'dynamical activity'~\cite{baie09}. Such
an approach has been demonstrated experimentally for a single driven colloidal
particle~\cite{blic07,gome09,mehl10}. The second strategy introduces a
multiplicative correction through an effective
temperature~\cite{cugl97a,kurc05} replacing $T$ in
Eq.~\eqref{eq:er}. Originally developed in the context of aging, glassy
dynamics and weakly driven systems, effective temperatures have also been
investigated in shear driven supercooled liquids~\cite{bert02a}.

Self-diffusion in sheared interacting suspensions has been studied extensively
in computer simulations~\cite{heye86,cumm91,rast96,foss00,bert02a} and
experiments~\cite{qiu88,bess07} as well as
analytically~\cite{indr95,morr96,szam04}.  A large body of publications
studies supercooled conditions in relation with the glass
transition~\cite{bert02a,bess07,krue10}. Most of these works focus on the
self-diffusion coefficient as this quantity is easily obtained from
experiments and simulations. The mobility of a tagged particle has been
addressed somewhat less prominently and mostly in analytic
calculations~\cite{blaw95,szam04}. In this Letter we determine numerically
both the full time-dependent velocity response and autocorrelation functions
for a tagged particle of a hard-core Yukawa suspension driven into a
nonequilibrium steady state through simple shear flow. In contrast to previous
work, explicit knowledge of the response function allows us to calculate and
discuss the mobility as a function of density and strain rate. We use a novel
method to efficiently obtain the time-dependent response function of the
velocity with respect to a small force applied to a single particle. Similar
methods to extract the response of a system using only unperturbed steady
state trajectories have been discussed in Refs.~\cite{chat04,bert07}.


\section{Sheared hard-core Yukawa fluid}

\begin{figure}[t]
  \centering
  \includegraphics[width=\linewidth]{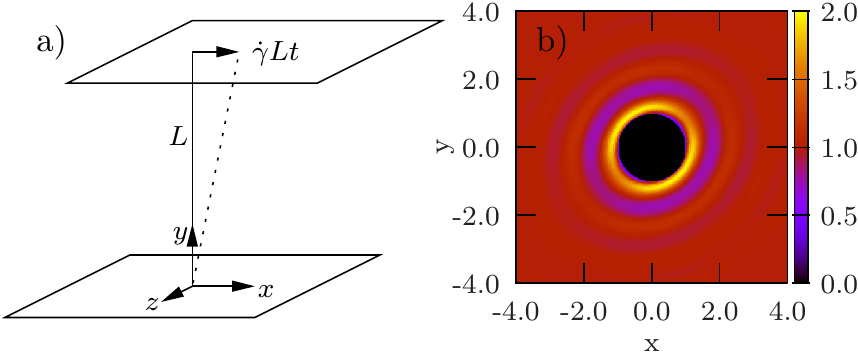}
  \caption{a) Simple shear flow with strain rate $\shr$. b) Pair distribution
    function $g(\x)$ in the $xy$ plane for volume fraction $\phi=0.3$ and
    strain rate $\shr=1$. Centered on any particle, the function $g(\x)$
    quantifies the probability to find another particle at $\x$. While
    isotropic in equilibrium, the pair distribution function is distorted
    through external flow.}
  \label{fig:flow}
\end{figure}

The $N$ colloidal particles interact through the purely repulsive Yukawa
potential
\begin{equation}
  \label{eq:yuk}
  u(r) =
  \begin{cases}
    \eps\frac{e^{-\kappa(r-1)}}{r} & (r\geqslant 1) \\
    \infty & (r<1)
  \end{cases}
\end{equation}
with hard core exclusion. The two potential parameters are the energy $\eps$
at contact and the screening length $\kap^{-1}$ determining the range of
interactions. Changing $\kap$ interpolates between hard-sphere (large $\kap$)
and coulombic (small $\kap$) interactions. Throughout the paper we employ
dimensionless units and measure length in units of the particle diameter $a$
and energies in units of $\kT$. The time scale $3\pi a^3\eta/\kT$ is set by
the time a particle diffuses a distance equal to its diameter. In particular,
employing these units the mobility and diffusion coefficient of a free
particle reduce to unity, $D_0=\mu_0=1$. We set $\kap^{-1}=0.2$ and choose
$\eps=8$ such that the liquid is stable for a large pressure
range~\cite{azha00}. We explore the liquid phase at four volume fractions
$\phi\equiv0.1,0.2,0.3,0.4$, where $\phi=\pi N(a/L)^3/6$ with $L$ the side
length of the cubic simulation box. The highest density $\phi=0.4$ is close to
the equilibrium freezing transition, which occurs at a pressure
$28.9$~\cite{azha00} (for $\phi=0.4$ the measured pressure in our simulation
is $27.6$). For comparison, the freezing transition in a hard sphere
suspension occurs at $\phi\simeq0.494$~\cite{ande02}.

We employ Brownian dynamics simulations, for details see the appendix. The
suspension is driven into a nonequilibrium steady state through simple shear
flow $\vec u(\x)=\shr y\vec e_x$ with strain rate $\shr$ (which equals the
P\'eclet number in our units), see Fig.~\ref{fig:flow}a). The equations of 
motion are $\dot\x_k=\vec v^0_k$ and
\begin{equation}
  \label{eq:lang}
  \dot{\vec v}^0_k = -\nabla_k U - [\vec v^0_k -\vec u(\x_k)] 
  + \vec f_k + \boldsymbol\xi_k,
\end{equation}
where the dimensionless mass is set to one. Physically, this choice implies
that momenta relax on the diffusive time scale. Besides the forces due to the
potential energy $U\equiv\sum_{k<k'}u(|\x_k-\x_{k'}|)$ we allow for direct
forces $\vec f_k$. The stochastic noise $\boldsymbol\xi_k$ modeling the
interactions with the solvent has zero mean and correlations
$\mean{\xi_{ki}(t)\xi_{k'j}(t')}=2\delta_{ij}\delta_{kk'}\delta(t-t')$, where
$i,j=x,y,z$ is the vector component. In Eq.~(\ref{eq:lang}), we neglect
hydrodynamic coupling between different particles due to the solvent.

For the shear flow turned on we correct the particle velocities $\{\vec
v^0_k\}$ by the external flow and investigate the relative velocity 
$\vec v_k=\vec v^0_k-\vec u(\x_k)$. We are interested in the dynamics 
of a single tagged particle interacting with the remaining $N-1$ particles 
in the suspension. Since all particles are identical we designate particle 1
as the tagged particle and drop the subscript; in the following $\x$ and 
$\vec v$ are the position and relative velocity of the tagged particle, 
respectively. We define the response of this velocity
\begin{equation}
  \label{eq:R}
  R_{ij}(t-t';\shr) \equiv \fd{\mean{v_i(t)}}{f_j(t')}
\end{equation}
with respect to an \textit{additional} small force $\vec f$. In addition, we
define the relative velocity autocorrelation matrix
\begin{equation}
  \label{eq:C}
  C_{ij}(t-t';\shr) \equiv \mean{v_i(t)v_j(t')}_0.
\end{equation}
The brackets $\mean{\cdots}_0$ refer to an average with respect to the
unperturbed steady state whereas $\mean{\cdot}$ is the average with the
external force applied. In equilibrium ($\shr=0$) the fluctuation-dissipation
theorem $R_{ij}(t)=C_{ij}(t)$ holds.


\section{Response function}

Sampling the correlation function~\eqref{eq:C} is straightforward. The direct
way to obtain the response matrix~\eqref{eq:R} from simulations would be
through a step perturbation of the force and the subsequent recording of the
tagged particle's velocity. Such a protocol has to be repeated many times and
the corresponding response function follows as the time-derivative of the mean
velocity. Here, we follow another route and determine the response function
through the path integral representation of the fluctuation-dissipation
theorem (FDT) for nonequilibrium steady states. The FDT in its general form
reads
\begin{equation*}
  R_{ij}(t-t';\shr) = \mean{ v_i(t) B_j(t')}_0
\end{equation*}
with observable $B_j(t)=\delta\ln P/\delta f_j(t)|_{\vec f=0}$ conjugate to
the perturbation force $\vec f$ acting on the tagged particle. The stochastic
path weight is
\begin{equation*}
  P[\{\nois_k(t)\};\vec f(t)] = \mathcal{N} \exp\left\{ -\frac{1}{4} 
    \sum_{k=1}^N\Int{t}\nois^2_k(t) \right\}
\end{equation*}
with normalization constant $\mathcal N$. From Eq.~\eqref{eq:lang} we see that
a perturbation of $\vec f$ is equivalent to a perturbation of $\nois$ with
$\delta\xi_i(t)/\delta f_j(t')=-\delta_{ij}\delta(t-t')$ we immediately obtain
$B_j=\xi_j/2$ and therefore
\begin{equation}
  \label{eq:R:nois}
  R_{ij}(t-t';\shr) = \frac{1}{2}\mean{v_i(t)\xi_j(t')}_0.
\end{equation}
Since in a computer simulation we have direct access to the noise we can
exploit such an expression to obtain the response function through a steady
state correlation function. While Eq.~\eqref{eq:R:nois} has been known
before~\cite{cala05,spec06}, to the best of our knowledge so far it has not
been exploited to obtain the response function numerically.

\begin{figure}[t]
  \centering
  \includegraphics[width=\linewidth]{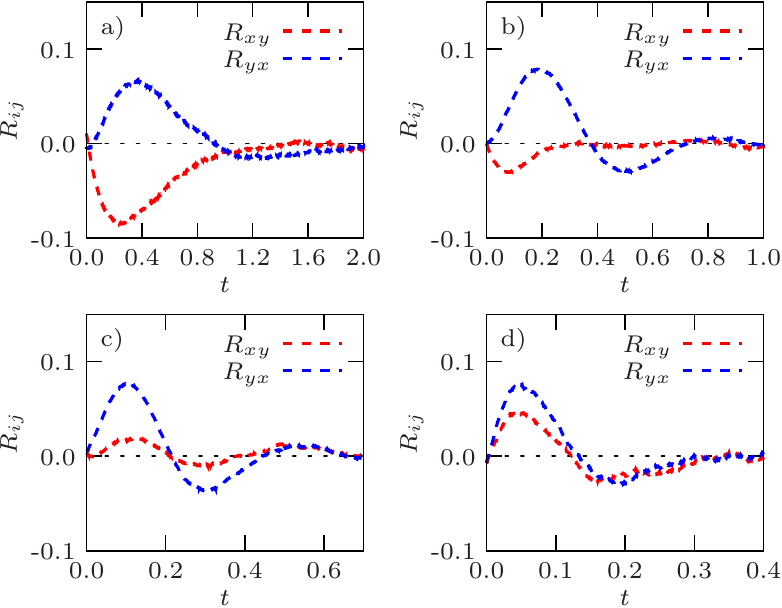}
  \caption{Comparison of the off-diagonal components $R_{xy}(t)$ and
    $R_{yx}(t)$ for strain rate $\shr=1.0$ and the four simulated volume
    fractions: a)~$\phi=0.1$, b)~$\phi=0.2$, c)~$\phi=0.3$, and
    d)~$\phi=0.4$. Increasing the density, the two curves approach each other
    until for the highest density they almost lie on top of each other. Note
    the changing time scale.}
  \label{fig:offdiag}
\end{figure}

To understand the influence of the shear flow on the particle motion it is
instructive to look at the off-diagonal components $R_{xy}$ and $R_{yx}$
plotted in Fig.~\ref{fig:offdiag}. The component $R_{yx}$ describes the mean
behavior of a tagged particle when we apply a force parallel to the shear flow
and measure its velocity perpendicular in $y$ direction. The behavior of
$R_{yx}$ can be explained by looking at the pair distribution function in
Fig.~\ref{fig:flow}b), which is deformed compared to its equilibrium isotropic
shape. In order to move faster at short times the particle moves up ($R_{yx}$
is positive) to overcome its neighbors through a region of lower probability
to encounter another particle. At a later time the particle is pushed back
($R_{yx}$ is negative) due to interactions with other particles which become
more pronounced at higher densities. Interchanging $x$ and $y$-direction, the
same arguments hold for the component $R_{xy}$. However, since we pull the
particle up it enters a region where the surrounding fluid moves faster due to
the shear flow. Hence, the particle is accelerated and the response of the
relative velocity is negative for small times. With increasing density the
particle cannot move far in $y$-direction, making the velocity differences
smaller. The qualitative difference between the two curves diminishes and for
$\phi=0.4$ both almost lie on top of each other. Hence, the effect of the
shear flow on a single particle is more and more symmetric as particle motion
becomes correlated at higher densities.


\section{Diffusion and mobility}

\begin{figure}[b!]
  \centering
  \includegraphics[width=\linewidth]{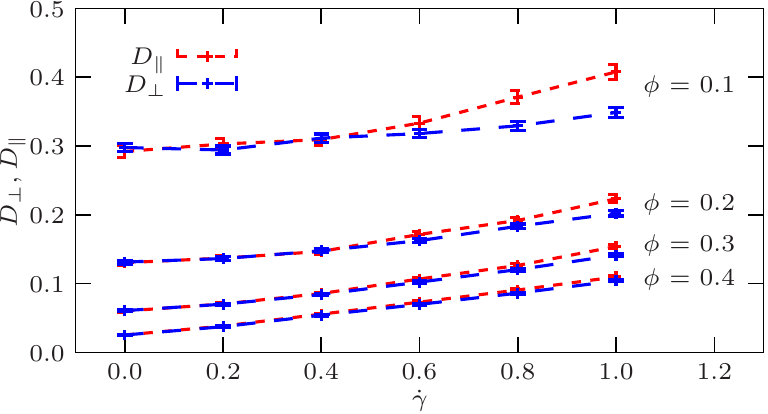}
  \caption{Diffusion coefficients $D_\parallel$ parallel and $D_\perp$
    perpendicular to the shear flow \textit{vs}. strain rate $\shr$ for the
    four different volume fractions $\phi$. For a free particle
    $D_\perp=D_\parallel=1$.}
  \label{fig:diff}
\end{figure}

We now turn to the nonequilibrium diffusion coefficients and mobilities,
\begin{equation}
  \label{eq:int}
  D_{ij} \equiv \IInt{t}{0}{\infty} C_{ij}(t), \quad
  \mu_{ij} \equiv \pd{\mean{v_i}}{f_j} = \IInt{t}{0}{\infty}R_{ij}(t),
\end{equation}
which are obtained through integrating the velocity autocorrelation and
response matrix, respectively. The mobility is defined as the velocity change
in response to a small force applied to the tagged particle, perturbing the
steady state reached through shearing the solvent. The diffusion coefficients
are related to the velocity autocorrelation through a Green-Kubo kind
relation. They are plotted in Fig.~\ref{fig:diff} and increase with increasing
strain rate. From our data we find that we can distinguish diffusion parallel
to the shear flow with $D_\parallel=D_{xx}$ and diffusion perpendicular with
$D_\perp=D_{yy}=D_{zz}$.  In shear flow the tagged particle moves between
layers of different flow velocity effectively leading to larger fluctuations,
allowing the particle to explore phase space faster. At low density the
diffusion $D_\parallel$ parallel to the shear flow is substantially enhanced
compared to $D_\perp$ even though we subtract out the external flow. However,
the difference between the two diffusion coefficients vanishes with increasing
density.

\begin{figure}[t]
  \centering
  \includegraphics[width=\linewidth]{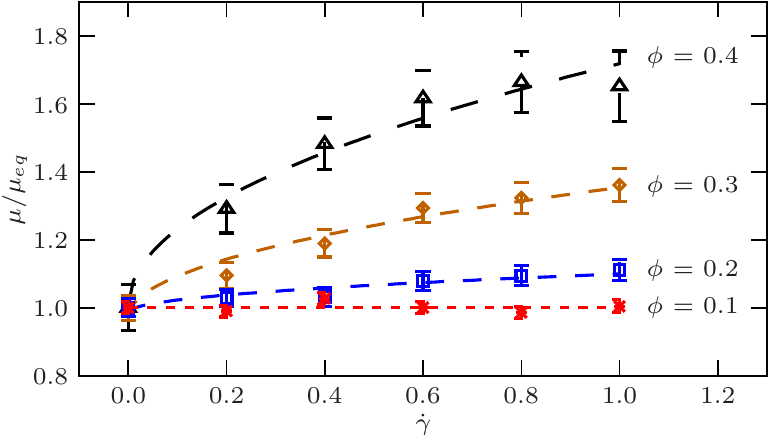}
  \caption{Reduced mobilities $\mu/\mu_\text{eq}$ \textit{vs}. strain rate
    $\shr$ for the four different volume fractions $\phi$. The lines are fits
    to Eq.~\eqref{eq:mu}.}
  \label{fig:mob}
\end{figure}

In Fig.~\ref{fig:mob} we plot the reduced mobility
$\mu(\phi,\shr)/\mu_\text{eq}(\phi)$ versus strain rate $\shr$ for the four
simulated densities with equilibrium ($\shr=0$) mobility $\mu_\text{eq}$. We
find that the diagonal components of the mobility matrix are equal within
error bars and we obtain the shown mobilities through averaging over the three
directions. The off-diagonal components are somewhat harder to obtain due to
large statistical errors but are clearly much smaller than their diagonal
counterparts (data not shown).  The dependence of the absolute value of the
mobility on the strain rate is rather weak and for the lowest density
$\phi=0.1$ it is even constant. Such a weak dependence suggests that the
ability of the solvent to reorganize in response to dragging the tagged
particle out of the 'cage' formed by neighboring particles is only slightly
affected by the presence of the shear flow. Going to supercooled conditions,
this is likely to break down~\cite{habd04}.

To explain the dependence of the mobility on the strain rate we consider the
mean velocity of the tagged particle from Eq.~(\ref{eq:lang}),
\begin{equation*}
  \mean{\vec v} = \mean{\vec F^{(1)}} + \vec f = -(N/L)^3\Int{\x}
  g(\x)\nabla u(\x) + \vec f.
\end{equation*}
Here, $g(\x;\phi,\shr,\vec f)$ is the pair distribution function to find a
second particle at $\x$ if there is a particle at the origin, see
Fig.~\ref{fig:flow}b), and $\vec F^{(1)}\equiv-\nabla_1U$ is the force exerted
by neighboring particles on the tagged particle. The effects of the shear flow
and the force $\vec f$ enter only through the structure information encoded in
the pair distribution function. We can expand $g$ into a Taylor series for
small forces $\vec f$. On the other hand, it is well known that the structure
of the suspension in the presence of shear flow is singularly perturbed from
its isotropic equilibrium form~\cite{dhont,russel}, requiring an asymptotic
expansion in powers of $\shr^{1/2}$. Such an expansion in lowest order leads
to
\begin{equation}
  \label{eq:mu}
  \mu(\phi,\shr)\approx\mu_\text{eq}(\phi)\left[1+\chi(\phi)\shr^{1/2}\right].
\end{equation}
In principle the coefficients $\chi(\phi)$ could be obtained from the
knowledge of the perturbed pair distribution function $g(\x)$. Here, we
determine them through fitting the mobility, see the lines in
Fig.~\ref{fig:mob}. The fits show a good agreement with the simulated data for
all strain rates and densities even though we have only retained the lowest
order of the expansion~(\ref{eq:mu}).


\section{Einstein relation}

\begin{figure*}[t]
  \centering
  \includegraphics[width=\linewidth]{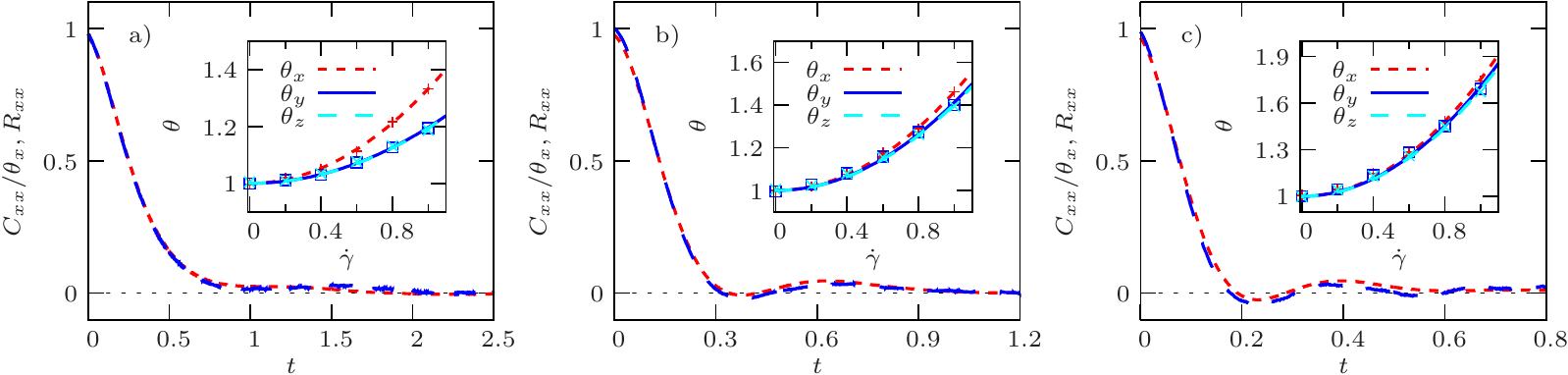}
  \caption{The response $R_{xx}(t)$ (dashed) and correlation $C_{xx}(t)$
    (dotted) functions for volume fractions a) $\phi=0.1$, b) $\phi=0.2$, and
    c) $\phi=0.3$. The correlation functions are scaled by a constant factor
    $1/\theta_x$ to match the initial decay of the response function. The
    insets show these factors as a function of strain rate for the three
    directions, where the lines are fits to Eq.~\eqref{eq:teff}.}
  \label{fig:teff}
\end{figure*}

In Fig.~\ref{fig:teff} we plot $R_{xx}(t)$ and $C_{xx}(t)$ as functions of
time for different volume fractions. The correlation functions have been
scaled by a constant factor $1/\theta_x$ to match the initial decay of the
response functions. This procedure reveals a rather good agreement between
response and correlation function even for longer times (this holds also for
the $yy$ and $zz$ components). We approximate the ratio
\begin{equation}
  \label{eq:teff}
  \frac{C_{ii}(t;\phi,\shr)}{R_{ii}(t;\phi,\shr)} \approx \theta_i(\phi,\shr)
  \approx 1 + \alpha_i(\phi)\shr^2
\end{equation}
by these constant factors. Using $R_{ii}(0)=1$ we can interpret
$\theta_i\approx\mean{v_i^2}_0$ as effective temperatures since they equal
approximately the velocity fluctuations of the tagged particle. We have also
checked the distribution functions of the velocity which are Gaussian with
width $\sqrt{\theta_i}$ as expected.

In Eq.~(\ref{eq:teff}) we expand the correlation function to second order in
the strain rate (due to the symmetry of simple shear flow there is no first
order) with fit parameters $\alpha_i$. This expansion becomes exact in the
case of interacting particles with linear forces~\cite{spec09}. In the insets
of Fig.~\ref{fig:teff} the factors $\theta_i$ for the three directions $x$,
$y$, and $z$ are shown to follow this quadratic prediction. Similar to the
diffusion coefficients we can distinguish a factor $\theta_\parallel=\theta_x$
parallel, and a factor $\theta_\perp=\theta_y=\theta_z$ perpendicular to the
shear flow. Again, increasing the density the difference between the two
directions vanishes.

Two points are noteworthy. First, no simple proportionality can be found
between the off-diagonal components of the response and the correlation
matrix. Both have a qualitatively different shape (data not shown).  In
particular, the off-diagonal response components are strictly zero at $t=0$
whereas, in nonequilibrium, the off-diagonal velocity correlations are
different from zero. Second, there is a fundamental difference compared to the
effective temperature discussed for non-stationary, glassy dynamics out of
equilibrium. There, fluctuation and dissipation are related by an effective
temperature at low frequencies (i.e., on long time scales) while the initial
decay of response and correlations is governed by the bath
temperature~\cite{cugl97a}. Moreover, the effective temperature evolves slowly
as the system approaches equilibrium. In contrast, in our case already the
initial decay is governed by a temperature $\theta_i>1$. The effect of this
temperature extends into the tails of response and correlation functions. It
is only for high densities that we observe a deviation in the tails as can be
seen in Fig.~\ref{fig:teff}c).

We can finally write down a simple generalized Einstein relation
\begin{equation}
  \label{eq:er:mod}
  D_{ii} = \theta_i\mu
\end{equation}
for the diagonal components of the diffusion matrix. Inserting the expansions
for mobility [Eq.~\eqref{eq:mu}] and effective temperature
[Eq.~\eqref{eq:teff}] we find that $D_{ii}-\mu_\text{eq}\sim\shr^{1/2}$ to
lowest order. Such a dependence was also found in molecular dynamics
simulations for a Lennard-Jones fluid~\cite{heye86,cumm91}. Due to the small
$\chi/\alpha_i$ ratios we cannot resolve this $\shr^{1/2}$ dependence here.

\begin{figure}[b!]
  \centering
  \includegraphics[width=\linewidth]{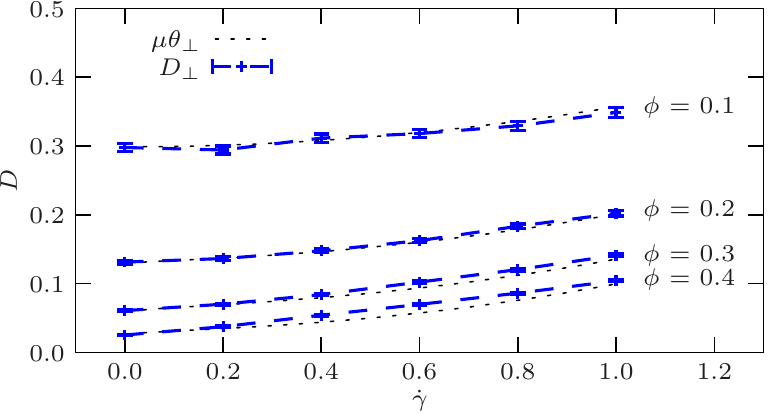}
  \caption{Test of the Einstein relation $D_\perp=\theta_\perp\mu$ with
    effective temperature $\theta_\perp$ from Eq.~(\ref{eq:teff}) and mobility
    from Eq.~(\ref{eq:mu}). The curves show a very good agreement for the
    lowest two densities but start to deviate at higher densities.}
  \label{fig:er}
\end{figure}

In Fig.~\ref{fig:er} we test the putative effective temperature by comparing
$\theta_\perp\mu$ to the numerically obtained diffusion coefficient $D_\perp$
perpendicular to the shear flow. The mobility for $\phi=0.1$ is independent of
strain rate and the diffusion follows the quadratic prediction
$D_\perp\propto\shr^2$. While we find a good agreement for the two lowest
densities, the effective temperature underestimates the diffusion coefficient
at intermediate strain rates and high densities since the diffusion
coefficient qualitatively changes and approaches a linear function
$D_\perp\propto\shr$ at high densities. This indicates that the differences in
the tails of response and velocity autocorrelation funtions become more
important. Also, higher order terms might be required in the expansion of
mobility and effective temperature.

\section{Experimental realization}

We briefly discuss how our findings could be tested in experiments. Of course,
the route via Eq.~\eqref{eq:R:nois} to obtain the velocity response matrix
through the explicit knowledge of the noise is not available in experiments.
Moreover, the direct route, i.e. perturbing only a single particle within a
suspension and measuring its time-dependent mean velocity, is experimentally
challenging and, as we find in our simulations, also statistically more
demanding.

Despite the difficulties it is still interesting to obtain this response
function since it immediately yields the nonequilibrium mobility. We now
assume that the tagged particle undergoes overdamped motion which is certainly
the relevant limit for experiments. The Langevin equation for the tagged
particle reads
\begin{equation}
  \label{eq:lang:od}
  \dot\x - \shr y\vec e_x = \vec F^{(1)} + \vec f + \nois.
\end{equation}
The replacement of the noise in Eq.~\eqref{eq:R:nois} by
Eq.~\eqref{eq:lang:od} is permissible since the Jacobian arising due to the
change of variables is independent of $\vec f$. We then obtain an
experimentally accessible expression for the response function
\begin{equation}
  \label{eq:ex}
  R_{ij}(t) = \frac{1}{2}\left[ C_{ij}(t)-\mean{v_i(t)F^{(1)}_j(0)}_0 \right]
\end{equation}
for components $i,j=y,z$ perpendicular to the shear flow. Let us assume that
we record the particle position $\x_k$ at times $t_k\equiv k\tau$ with time
resolution $\tau$, e.g., through video microscopy. The velocity is then
approximated through the finite difference $\vec v_k=(\x_k-\x_{k-1})/\tau$. In
principle, the force $\vec F^{(1)}$ on the tagged particle can be calculated
from the knowledge of the pair potential and the positions of all neighboring
particles.


\section{Conclusions}

We have studied a hard-core Yukawa colloidal system at different densities
driven into a nonequilibrium steady state through shear flow. In particular,
we investigate diffusion and mobility of a single tagged particle for four
volume fractions $\phi$ and intermediate strain rates $\shr\leqslant1$. The
self-diffusion coefficient is calculated through the Green-Kubo relation from
the single particle's velocity autocorrelation function. The mobility is
obtained from the particle's response function through integration. For
systems governed by stochastic dynamics, this response function can be
obtained efficiently from the correlation function Eq.~(\ref{eq:R:nois})
measured in the unperturbed steady state. While for low densities we can
clearly distinguish quantities measured parallel and perpendicular to the shear
flow, this difference vanishes for high densities.

Surprisingly, the diagonal components of the response (i.e., the response is
measured in the direction of the applied force) and correlation matrix can be
matched over a large time range. The resulting proportionality factor can be
interpreted as an effective temperature, effectively restoring the Einstein
relation connecting diffusion and mobility. Moreover, this proportionality
factor is well approximated by a quadratic expansion in the strain rate. It
will be important to study how general such a simple effective temperature is
for driven interacting colloidal suspensions and whether it extends to other
observables. We believe that the methodology presented here will lead to new
insights in the numerical study of dense colloidal suspension, e.g., for
microscopic stress fluctuations~\cite{spec09,zaus09}. Finally, the influence
of hydrodynamic interactions on our results remains to be investigated.


\acknowledgments

TS gratefully acknowledges financial support by the Alexander-von-Humboldt
foundation and by the Director, Office of Science, Office of Basic Energy
Sciences, Materials Sciences and Engineering Division and Chemical Sciences,
Geosciences, and Biosciences Division of the U.S. Department of Energy under
Contract No.~DE-AC02-05CH11231. Financial support by the DFG through SE 1119/3
is also acknowledged.


\section{Appendix: Simulation details}

The simulated systems consist of $N=1728$ particles in a cubic simulation
box. Since we are interested in the bulk behavior of the suspension we employ
periodic boundary conditions and account for the shear flow through
Lees-Edwards sliding bricks. The equations of motion are integrated by a
stochastic version of the velocity Verlet algorithm~\cite{frenkelsmit}, where
the velocity appearing in the force term at the right hand side of
Eq.~\eqref{eq:lang} is taken from the mid-step velocity. The time step is set
to $\Delta t=0.0005\ll(\kappa\sqrt{\mean{v_i^2}})^{-1}\sim 0.08\dots0.2$. We
equilibrate the suspension and then slowly increase the strain rate to the
final value. Correlation functions have been obtained by averaging over 400
particle trajectories with 300,000 time steps each. These trajectories have
been determined in two independent runs from randomly chosen particles.

To implement the hard-core repulsion and prevent particles from overlapping,
the following simple algorithm is employed (see also
Refs.~\cite{stra99,foss00} and references therein). After every particle has
been moved, but before new forces are calculated, we store all overlapping
particle pairs and remove these overlaps as follows. For each pair both of the
particles are moved backwards in time along their respective velocity vector
up to the point where they collided. This time $0 < s < \Delta t$ is
stored. Knowing the positions and velocities at the impact, we compute the
connection vector $\vec e$ between both particles. We decompose the velocities
into $\vec{v}_{1,2}^{\parallel}=\vec{e}\vec{e}^T\vec{v}_{1,2}$ and
$\vec{v}_{1,2}^{\perp}=(\id-\vec{e}\vec{e}^T)\vec{v}_{1,2}$. Only the parts
parallel to $\vec{e}$ can change. Using the usual elastic collision rule
preserving momentum and kinetic energy of the particles we obtain the
after-collision velocities
$\vec{v}_1'=\vec{v}_1^{\perp}+\vec{v}_2^{\parallel}$ and
$\vec{v}_2'=\vec{v}_2^{\perp}+\vec{v}_1^{\parallel}$. From the positions of
their collision the particles are then propagated forward with time step $s$
along the new velocity vectors. This procedure is repeated as long as
overlapping pairs exist.



\end{document}